\documentclass[aps,prx,twocolumn,showpacs,superscriptaddress,longbibliography]{revtex4-2}

\usepackage[english]{babel}
\usepackage[utf8]{inputenc}
\usepackage{graphicx}
\usepackage{braket}
\usepackage[most]{tcolorbox}
\usepackage{mathrsfs}
\usepackage{bbold}
\usepackage{subcaption}
\usepackage[unicode=true,pdfusetitle,
 bookmarks=true,bookmarksnumbered=false,bookmarksopen=false,
 breaklinks=false,pdfborder={0 0 0},pdfborderstyle={},backref=false,colorlinks=true]
 {hyperref}
 \hypersetup{
 citecolor=blue,
 urlcolor=blue,
 linkcolor=blue}

 \usepackage{latexsym,amsmath,verbatim}
\usepackage{amssymb} 
\usepackage{amsfonts}
\usepackage{colortbl}
\usepackage{array}
\usepackage{stackrel}
\usepackage{bm}
\usepackage{nicefrac}
\usepackage{rotating}
\usepackage{float}
\usepackage[final]{pdfpages}
\usepackage{lipsum}
\usepackage{soul}

\makeatletter
\AtBeginDocument{\let\LS@rot\@undefined}
\makeatother

\begin{document}
\title{Synchronization, Kinematic Waves and Spike-Phase-Separation in Feedback Ising Neural Networks on Heterogeneous Graphs}
\author{Anna Poggialini}
\affiliation{Department of Biomedical Sciences, University of Padova, Padova, Italy}

\author{Irem Topal}
\affiliation{Department of Biomedical Sciences, University of Padova, Padova, Italy}

\author{Fabrizio Lombardi}
\affiliation{Department of Biomedical Sciences, University of Padova, Padova, Italy}

\author{Daniele De Martino}
\affiliation{Biofisika Institute/Fundacion Biofisika Bizkaia (CSIC, UPV-EHU) and Ikerbasque Foundation, Bilbao 48013, Spain}

\begin{abstract}
Structural heterogeneity constrains collective dynamics in complex systems.  However, its analytical tractability out of equilibrium remains limited. In this work, we study a class of kinetic Ising neural networks driven out of equilibrium by a homeostatic feedback loop between the  neuronal excitability  and the population  firing rate. Using a Curie-Weiss heterogeneous mean-field approximation validated by Monte Carlo simulations, we provide an analytical characterization of how a macroscopic synchronized limit cycle emerges via an Andronov-Hopf bifurcation on heterogeneous networks. We derive closed-form phase boundaries  and show that the onset of oscillations is explicitly controlled by network heterogeneity through the degree moment ratio $\alpha = \langle k^2 \rangle / \langle k \rangle$.  Degree heterogeneity decouples the spiking rate per neuron $m$ from the spiking rate per synapse $u$, generating  physical phenomena absent in homogeneous systems. These include (i) kinematic waves of sequential, degree-ordered activations propagating from the network periphery to the hubs, and (ii) a low-temperature phase-separated state emerging via a pitchfork bifurcation. We prove that for highly heterogeneous topologies, this phase-separated fixed point stabilizes and dynamically destroys the synchronized limit cycle. These results provide a mathematical framework for understanding how heterogeneity  regulates macroscopic oscillations and out-of-equilibrium transitions in neural networks.  
\end{abstract}
\maketitle

\section{Introduction}

In living systems, structure  and  function are intimately connected. While structure is subject to changes to address functional needs, structural constraints may in turn shape function,    in a constant evolutionary cross-talk that underlies self-organization in biological and ecological systems. As a result, the emergent structure is  highly heterogeneous. 

With their complex multiscale organization,   biological neural networks are a prominent example of this structure-function interplay. 
Their units, neuronal cells, are strongly heterogeneous, with the distribution of proteins  on the neuron's membrane and dendrites that strongly influence neurons function. 
At the same time, the large heterogeneity in neuronal structures originates  from adaption and functional specialization.  These different types of neurons are  the building blocks of large-scale networks that exhibit a wide variety of  collective behaviors. However, how and to what extent the structure of the network constrains (or influences)  collective modes and state transitions  in biological networks remains poorly understood~\cite{Bullmore2009,Sporns2011}.

Statistical mechanics provides a natural framework for studying how structural heterogeneity influences collective behavior, treating the variability of local environments as a form of quenched or annealed disorder whose consequences can be analyzed systematically. In this context, classical  neural network models were formulated at equilibrium and focused primarily on storage capacity and associative memory~\cite{Hopfield1982,Amit1985}. However, real neuronal networks operate out of equilibrium and display  a wide range of spontaneous oscillatory activity that spans multiple spatial and temporal scales~\cite{Buzsaki2006}.  
Models based on excitatory--inhibitory (EI) balance~\cite{vanVreeswijk1996,Brunel2000,borgers2005ei,wilsoncowan1972} or coupled oscillators~\cite{Kuramoto1984,Acebron2005,Buendia_2022_sync} can capture these oscillations, but their analytical tractability on heterogeneous graphs remains limited.

A complementary approach, recently introduced in \cite{de2019feedback, de2019oscillations, sinelshchikov2023emergence}, equips Ising-type spin models with a linear feedback between the order parameter (magnetization) and the control parameter (external field).
On fully connected graphs and regular lattices, this mechanism replaces the equilibrium ferromagnetic transition with an out-of-equilibrium synchronization transition: the paramagnetic fixed point loses stability via a supercritical Andronov--Hopf bifurcation, giving rise to a macroscopic limit cycle~\cite{de2019feedback, de2019oscillations, sinelshchikov2023emergence}.
This approach opens the possibility of classifying specific brain oscillatory patterns in terms of bifurcation types, in analogy with the universality classes of equilibrium critical phenomena.
When inferred from magneto-encephalography (MEG) recordings of the resting human brain, the feedback Ising model captures the coexistence of scale-free neuronal avalanches and scale-specific oscillations, placing the operating regime of the brain near a non-equilibrium critical point~\cite{lombardi2023statistical}. 

This and related analyses were based on the assumption of homogeneous connectivity. However, real neuronal circuits are far from homogeneous: even within a single cell type, neurons exhibit graded variability in their morphological, electrophysiological, and transcriptomic properties ~\cite{Dahmen2025}---a form of ``within-type'' heterogeneity that amounts to quenched disorder 
and has profound consequences for network dynamics, computation, and self-organization. Understanding how such heterogeneity shapes the collective behaviors  
is therefore an essential step toward a realistic theory.


Here, we investigate the class of feedback Ising models on random graphs with heterogeneous degree distributions, which are widely recognized as essential ingredients in the modeling of real-world networks~\cite{Albert2002,Newman2003}.
%
We develop a Curie--Weiss (heterogeneous mean-field) theory and show that degree heterogeneity promotes the emergence of two distinct order parameters: 
the firing rate \emph{per neuron}, $m$, and the firing rate  \emph{per synapse}, $u$, whose dynamical interplay is absent  in the homogeneous case and governs the phase diagram.
%
We derive analytical bifurcation lines in closed form as functions of the feedback strength and the first two moments of the degree distribution, and validate them against Monte Carlo simulations on Erd\H{o}s--R\'enyi, scale-free, and log-normal graphs.

The analysis uncovers two collective phenomena that are absent in the homogeneous case: (i) \emph{kinematic waves}, degree-ordered sequential activations in which low-degree nodes respond first to perturbations and high-degree nodes follow with a delay, producing a wave-like propagation from the network periphery toward the hubs; (ii) \emph{phase-separated fixed point} with $m=0$ but $u\neq 0$, which  emerges at low temperatures via a supercritical pitchfork bifurcation and where degree classes split into two groups with opposite magnetizations.

The remainder of the paper is organized as follows. In Sec.~II we present the model and simulation results, develop the Curie--Weiss theory, derive the Hopf bifurcation lines and phase diagrams, and analyze the kinematic waves and the phase-separated state.
Sec.~III summarizes our conclusions and outlines directions for future work.
Technical details of the system-size expansion are collected in the Appendix.
\section{Results}

\begin{figure*}[t]
    \centering
    \includegraphics[width=\textwidth]{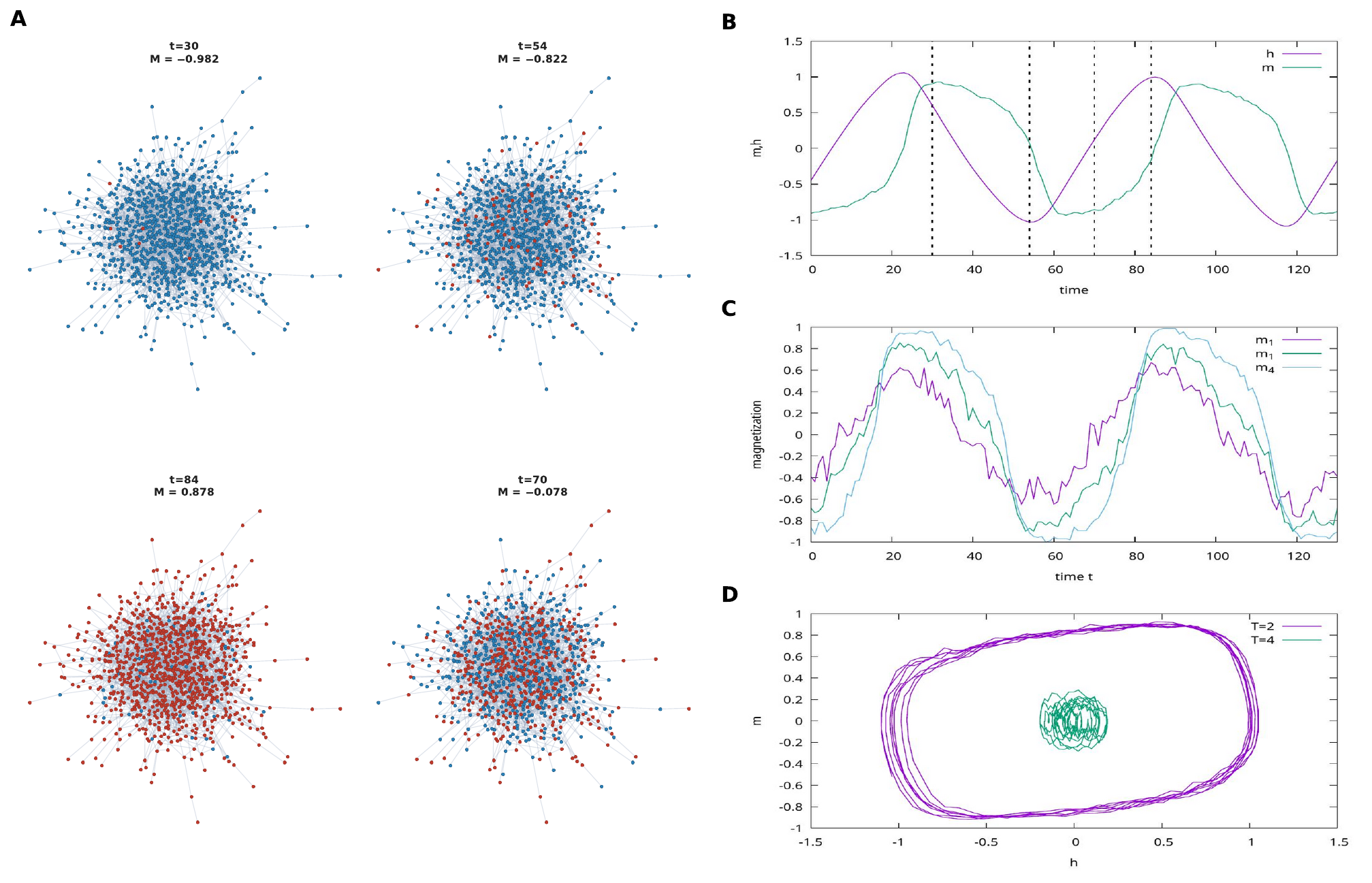}
    \caption{{\bf Feedback Ising model on a heterogeneous network.}
    {\bf A.}~Microscopic spin configurations ($s_i=+1$ in blue, $s_i=-1$ in red) at four representative times along the limit cycle.
    {\bf B.}~Time series of the global magnetization $m(t)$ and the feedback field $h(t)$, showing nonlinear self-sustained oscillations. Dashed lines mark the snapshots of panel~A.
    {\bf C.}~Degree-resolved magnetizations $m_k(t)$ for different degree classes, illustrating heterogeneous yet globally synchronized oscillatory dynamics.
    {\bf D.}~Phase-plane trajectory in the $(m,h)$ plane \textcolor{black}{below ($T=2$) and above the critical temperature ($T=4$). At $T=4$  oscillations are suppressed and the system relaxes to the paramagnetic fixed point.} \textcolor{black}{Simulation parameters: $N=1000$ nodes, Erd\H{o}s--R\'enyi graph with average degree $z=4$ (1-core), inverse temperature $\beta=0.5$, feedback strength $c=0.1$.}
    }
    \label{fig:panel}
\end{figure*}

We consider a kinetic Ising model defined on a general graph.
Spin variables $s_i=\pm 1$ are associated with the nodes of the graph, and their interactions are governed by the Hamiltonian
\begin{equation}
    H(\vec{s}) = - J \sum_{\langle i,j\rangle}  s_i s_j 
        -h \sum_{i} s_i \,,
    \label{eq:Hamiltonian}
\end{equation}
where the first sum runs over all edges $\langle i,j\rangle$ of the graph and $h$ denotes a uniform external field.
Each spin is updated asynchronously via Glauber dynamics with spin flip rates
\begin{eqnarray}
  w_i(\boldsymbol{s})=\frac{1}{2}\Big[1 - s_i \tanh\big(\beta\, H_i^{\rm eff}(\boldsymbol{s})\big)\Big]\, \\
H_i^{\rm eff}(\boldsymbol{s}) = \sum_{j} A_{ij}\, s_j + h,
\end{eqnarray}
where $H_i^{\rm eff}$ is the effective local field acting on the spin $i$ and $A_{ij}$ is the adjacency matrix of the underlying graph, i.e. $A_{ij}=1$ if the nodes are connected by a link, $A_{ij}=0$ otherwise. We consider undirected graphs, and the matrix $A$ is symmetric.
The rates satisfy detailed balance with respect to the Boltzmann--Gibbs distribution $p(\vec{s})\propto e^{-\beta H}$, so that the unperturbed dynamics relaxes to thermal equilibrium.
We drive the system out of equilibrium by coupling the external field $h$ to the instantaneous magnetization $m=\frac{1}{N} \sum_i s_i$ through a linear feedback rule governed by a single parameter~$c>0$:
\begin{equation}
\Delta h = -c\, m\, \Delta t\,,   
\label{eq:feedback}
\end{equation}
with $\Delta t =1/N$, so that the feedback is updated after each single-spin flip \footnote{Alternative choices of the feedback timescale affect the entropy production rate and depend on the amount of information available to implement the control loop (here assumed to be complete), but lead to qualitatively similar dynamics.}.
On fully connected graphs and on regular lattices, this feedback mechanism replaces the equilibrium ferromagnetic transition with an out-of-equilibrium synchronization transition: the paramagnetic fixed point loses stability via a supercritical Andronov--Hopf bifurcation, giving rise to a macroscopic limit cycle~\cite{sinelshchikov2023emergence}.

Here, we investigate how this scenario is modified on heterogeneous networks.
Figure~\ref{fig:panel} summarizes simulation results for an Erd\H{o}s--R\'enyi (ER) graph with average degree $z=4$.
At feedback strength $c=0.1$ and temperature $T=2$, far below the equilibrium critical temperature of the Ising model on this graph,  the system exhibits self-sustained oscillations (Fig.~\ref{fig:panel}A, B). 
Resolving the dynamics by degree class reveals that different degree groups oscillate with distinct amplitudes and phases, yet remain globally synchronized (Fig.~\ref{fig:panel}C).
Raising the temperature above $T_c$ ($T=4$) completely suppresses  oscillations: the phase-plane trajectory spirals inward to the paramagnetic fixed point (Fig.~\ref{fig:panel}D).
These observations raise a natural question: how do the onset, shape, and stability of the limit cycle depend on the network topology and, in particular, on the degree of heterogeneity?
To address this, in the next section we develop a mean-field theory for random graphs that yields analytical phase diagrams.

\subsection{Curie--Weiss approximation and analytical phase diagrams}
\label{sec:CW}

\subsubsection{Heterogeneous mean-field equations}

In the heterogeneous mean-field (HMF) approximation, all nodes sharing the same degree~$k$ are assumed to have the same average magnetization~$m_k$.
The effective field experienced by a degree-$k$ node depends on the \emph{neighbor magnetization}~$u$, defined as the mean magnetization encountered when following a randomly chosen edge.
Because an edge leads to a degree-$k$ node with probability $k\,p_k/\langle k\rangle$, one has
\begin{equation}
    u = \sum_k \frac{k\, p_k}{\langle k\rangle}\, m_k \,.
    \label{eq:u_def}
\end{equation}

At equilibrium, requiring consistency between the local fields and the thermal expectation values yields the standard self-consistency equations for the Ising model on random graphs ~\cite{leone2002ferromagnetic}:
\begin{align}
    m_k^* &= \tanh\!\left(\beta k\, u^* + \beta h\right), \label{eq:mf_mkstar}\\
    u^* &= \sum_k \frac{k\, p_k}{\langle k\rangle}\, m_k^* \,. \label{eq:mf_ustar}
\end{align}
These reduce to the classical Curie--Weiss equation in the homogeneous (fully connected) limit.

To study stability and out-of-equilibrium effects, we follow the approach of Ref.~\cite{sinelshchikov2023emergence} and promote $m_k$, $u$, and $h$ to dynamical variables.
Each degree class relaxes toward its instantaneous mean-field expectation,
\begin{equation}
    \dot m_k = -m_k + \tanh\!\big(\beta k\, u + \beta h\big)\,, \label{eq:dyn_mk}
\end{equation}
and the neighbor magnetization $u(t)$ evolves consistently with its definition~\eqref{eq:u_def}.
The key ingredient driving the system out of equilibrium is the feedback on the external field,
\begin{equation}
    \dot h = -c\,\sum_k p_k\, m_k = -c\,m\,,
    \label{eq:dyn_h}
\end{equation}
which couples the control parameter $h(t)$ to the instantaneous global magnetization, $m = \sum_k p_k\,m_k$.
The complete mean-field dynamics then reads
\begin{equation}
   \left\{
   \begin{aligned}
        &\dot m_k = -m_k + \tanh\!\big(\beta k\, u +\beta h\big),\\
        &\dot u    = -u + \sum_k \frac{k\, p_k}{\langle k\rangle}\,\tanh\!\big(\beta k\, u + \beta h\big),\\
        &\dot h    = -c\,m\,.
    \end{aligned}
    \right.
    \label{Eq.systEq}
\end{equation}
Note that the equation for $\dot{u}$ is obtained by substituting Eq.~\eqref{eq:dyn_mk} into the time derivative of Eq.~\eqref{eq:u_def}.
The complete and detailed derivation of system~\eqref{Eq.systEq}  is reported in Appendix \ref{App.VanKampen}.
Two features distinguish this system from its homogeneous counterpart.
First, the global magnetization $m = \sum_k p_k\,m_k$ and the neighbor magnetization $u = \sum_k (k\,p_k/\langle k\rangle)\,m_k$ are now \emph{distinct} order parameters: $m$ is a per-neuron average, while $u$ is a per-edge average weighted toward high-degree nodes.
On a regular graph the two coincide, but on a heterogeneous network their interplay governs the dynamical phase diagram and gives rise to several nontrivial phenomena, as discussed below.

\subsubsection{Linear stability analysis and bifurcation condition}

The paramagnetic state $m_k = u = h = 0$ is always a fixed point of Eqs.~\eqref{Eq.systEq}. 
To determine the onset of oscillations, we linearize Eqs.~\eqref{Eq.systEq} around this fixed point. 
Writing $m = \sum_k p_k\,m_k$ for the global magnetization perturbation, the linearized system in the variables $(m, u, h)$ reads
\begin{align}
\dot{m} &= -m+\beta\!\left(\langle k \rangle\, u+h\right), \label{eq:lin_m}\\  
\dot{u} &= -u+\beta\!\left(\frac{\langle k^2 \rangle}{\langle k \rangle}\, u+h\right), \label{eq:lin_u}\\
\dot{h} &= -c\,m\,. \label{eq:lin_h}
\end{align}
The characteristic polynomial of the Jacobian is
\begin{equation}
\lambda^3 + a\,\lambda^2 + b\,\lambda + d=0\,,
\label{eq:charpoly}
\end{equation}
with coefficients
\begin{equation}
    \begin{aligned}
        a &= 2-\beta\,\frac{\langle k^2 \rangle}{ \langle k \rangle}\,, \\
        b &= 1 + \beta\!\left(c-\frac{\langle k^2 \rangle}{ \langle k \rangle}\right), \\
        d &= \beta c\!\left(1 + \beta \langle k \rangle -\beta\, \frac{\langle k^2 \rangle}{ \langle k \rangle}\right).
    \end{aligned}
    \label{eq:RH_coeffs}
\end{equation}
By the Routh--Hurwitz criterion, the fixed point is stable if and only if
\begin{equation}
a>0\,, \quad d>0\,, \quad a\,b -d >0\,.
\label{eq:RH_conditions}
\end{equation}
Inequalities \ref{eq:RH_conditions} imply $b > 0$ when $c>0$.

It is instructive to examine  each condition separately. Introducing for brevity the moment ratio $\alpha \equiv \langle k^2\rangle/\langle k\rangle$, the first inequality yields $\beta < 2/\alpha$, while the second gives $\beta < 1$ (always weaker than the first for any physical graph). The third inequality is the most interesting: expanding $a\,b-d$ as a quadratic inequality  in $\beta$ one finds
\begin{equation}
\bigl(\alpha^2 - c(\alpha-1)\bigr)\beta^2 + (c-3\alpha)\beta + 2 > 0\,.
\label{eq:abd_quadratic}
\end{equation}
Remarkably, the discriminant of the associated quadratic equation simplifies to
\begin{equation}
\Delta = (c-3\alpha)^2 - 8\bigl(\alpha^2 - c(\alpha-1)\bigr) = (c+\alpha)^2 - 8c\,,
\end{equation}
and the smallest positive root, which defines the bifurcation threshold, reads in closed form
\begin{equation}
\boxed{
\beta_c(c) = \frac{3\alpha - c - \sqrt{(c+\alpha)^2 - 8c}}{2\bigl(\alpha^2 - c(\alpha-1)\bigr)}\,.
}
\label{eq:betac_new}
\end{equation}
Numerical inspection across a wide variety of degree distributions (regular, Erd\H{o}s--R\'enyi, heterogeneous and heavy-tailed) shows that the condition $ab-d>0$ is \emph{always} more restrictive than both $a>0$ and $d>0$ in the physically relevant range $c>0$, and thus  Eq.~\eqref{eq:betac_new} alone determines the Hopf-bifurcation line.

Two asymptotic limits of Eq.~\eqref{eq:betac_new} are particularly illuminating:
\begin{equation}
\beta_c(0) = \frac{1}{\alpha} = \frac{\langle k\rangle}{\langle k^2\rangle}\,, \qquad 
\beta_c(\infty) = \frac{1}{\alpha-1}\,,
\label{eq:betac_limits}
\end{equation}
corresponding to absence of feedback ($c=0$)  and strong feedback ($c \to \infty$), respectively. For vanishing feedback, $\beta_c = \langle k\rangle/\langle k^2\rangle$ coincides with the classical mean-field result for the ferromagnetic transition on random graphs~\cite{leone2002ferromagnetic}, as expected in the absence of feedback.
For strong feedback, the critical temperature saturates at $\beta_c = 1/(\alpha-1)$, which for heterogeneous networks (large~$\alpha$) is only marginally different from the value expected for $c\to 0$. 
The full range between the two limits is
\begin{equation}
\frac{\beta_c(\infty)}{\beta_c(0)} = \frac{\alpha}{\alpha-1}\,,
\end{equation}
so that broad degree distributions ($\alpha\gg 1$) are essentially insensitive to feedback strength: the network topology, and not the feedback, controls the onset of oscillations.
Conversely, on homogeneous graphs ($\alpha$ close to its minimum value $z$), feedback has the strongest effect, with $\beta_c$ spanning a sizeable range when $c$ varies.
Crucially, the $c$-dependence of $\beta_c$ arises entirely through the $a\,b-d$ condition. In marked contrast to the lattice case where the Hopf threshold is independent of $c$~\cite{sinelshchikov2023emergence}, here the interplay between the two order parameters $m$ and $u$ makes the bifurcation sensitive to the control gain. 
Note also that the combination $\alpha=\langle k^2\rangle/\langle k\rangle$,  which controls both limits, coincides with the moment ratio that sets the classical ferromagnetic transition: broad degree distributions with large second moment lower $\beta_c$, facilitating the onset of oscillations---a network-topology effect with no counterpart on regular lattices.

\subsubsection{Application to Erd\H{o}s--R\'enyi graphs}

For a Poissonian (Erd\H{o}s--R\'enyi) graph with mean degree $z$, one has $\langle k \rangle=z$, $\langle k^2 \rangle=z(z+1)$ and therefore $\alpha=z+1$.
Substituting this into Eq.~\eqref{eq:betac_new} gives the explicit form
\begin{equation}
\beta_c(c) = \frac{3(z+1) - c - \sqrt{(c+z+1)^2 - 8c}}{2\bigl((z+1)^2 - c\,z\bigr)}\,,
\label{eq:betac_ER}
\end{equation}
with limits 
\begin{equation}
\beta_c(0) = \frac{1}{z+1}\,, \qquad \beta_c(\infty) = \frac{1}{z}\,.
\end{equation}

\begin{figure*}[t]
    \centering
    \includegraphics[width=\textwidth]{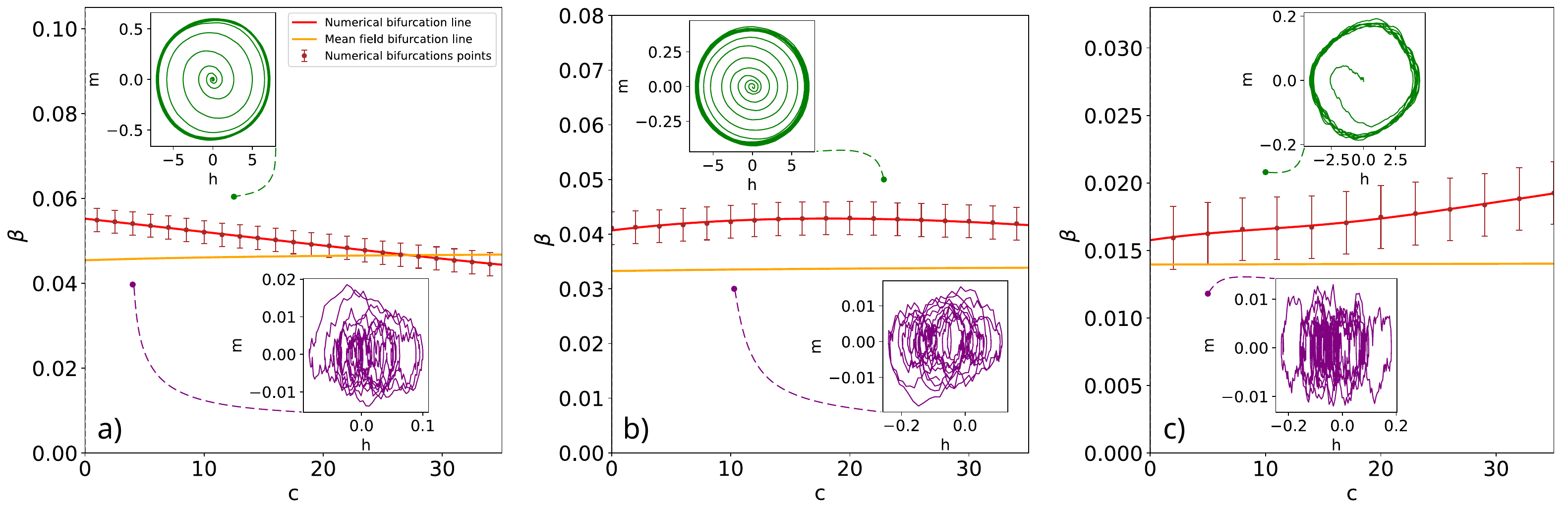}
    \caption{{\bf Analytical phase diagrams in the $(c,\beta)$ plane from the Curie--Weiss linear stability analysis.}
    \textit{Left:} Erd\H{o}s--R\'enyi graph with $z=20$.
    \textit{Center:} Scale-free network with exponent $\gamma = 2.1$ and $k_{\min}=20$.
    \textit{Right:} Log-normal degree distribution with $\langle k\rangle=25$ and $\sigma=1$ ($\mu = \ln\langle k\rangle-\sigma^2/2$).
    \textit{Insets:} Phase-plane trajectories $(m,h)$ from Monte~Carlo simulations confirm the analytical predictions.}
    \label{fig:phase_diagrams}
\end{figure*}

\textcolor{black}{For $z=20$ used in our numerical experiments to make  Erd\H{o}s--R\'enyi results comparable to those obtained 
on scale-free and log-normal graphs with a similar mean degree},  
$\beta_c$ smoothly and monotonically interpolates between 
$1/21\approx 0.048$ at $c=0$ and $1/20=0.050$ at $c\to\infty$, a variation of less than $5\%$ over the entire range of feedback strengths (Fig.~\ref{fig:phase_diagrams}). This near-insensitivity to $c$ is a direct consequence of 
Eq.~\eqref{eq:betac_limits}: for large $z$, \textcolor{black}{we have 
$\alpha/(\alpha-1)=(z+1)/z\to 1$, and the  topology completely  dominates over feedback in setting the bifurcation threshold}. 
Physically, strong feedback hinders the buildup of ferromagnetic order by rapidly adjusting the field against the magnetization, thereby pushing the bifurcation to lower temperatures 
(higher $\beta_c$); however, unlike what one might naively expect, this effect saturates and the Hopf threshold remains bounded, its $c$-dependence becoming negligible for dense graphs.


\subsubsection{Extension to strongly heterogeneous networks: power law and log-normal)}

Next, we consider graphs with power-law degree distributions $p(k) \propto k^{-\gamma}$ with $k \geq k_{\min}$, where the maximum degree is set by a structural cutoff $k_{\max}\sim N^{1/2}$ (or, alternatively, the natural cutoff $k_{\max}\sim N^{1/(\gamma-1)}$).
The relevant moments are
\begin{equation}
\langle k \rangle = 
\begin{cases}
\displaystyle \frac{\gamma-1}{\gamma-2}\,k_{\min}, & \gamma > 2, \\[6pt]
\sim\!\log N, &\gamma = 2,
\end{cases}
\label{eq:SF_k1}
\end{equation}
\begin{equation}
\langle k^2 \rangle = 
\begin{cases}
\displaystyle \frac{\gamma-1}{\gamma-3}\,k_{\min}^2, & \gamma > 3, \\[6pt]
\sim\!\log N, &\gamma = 3,\\[6pt]
\sim\!N^{(3-\gamma)/2}, & 2<\gamma < 3\;\text{(structural cutoff)}.
\end{cases}
\label{eq:SF_k2}
\end{equation}
For $\gamma > 3$, both moments are finite and the moment ratio $\alpha=\langle k^2\rangle/\langle k\rangle$ reads
\begin{equation}
\alpha = \frac{\gamma-2}{\gamma-3}\,k_{\min}\,,\qquad \gamma>3\,,
\end{equation}
which can be substituted directly into Eq.~\eqref{eq:betac_new} to obtain the full $\beta_c(c)$ curve for scale-free graphs.

For $2 < \gamma \leq 3$, the second moment diverges with system size ($\langle k^2\rangle \sim N^{(3-\gamma)/2}$ with the structural cutoff), so that $\alpha\to\infty$ and $\beta_c \to 0$ as $N\to\infty$ and the system is always in the oscillatory phase at any nonzero $\beta$ in the thermodynamic limit.
At finite $N$, replacing $\langle k^2\rangle$ by its $N$-dependent expression in~\eqref{eq:betac_new} yields finite-size estimates that decay as power laws in~$N$, i.e. 
\begin{equation}
    \beta_c(c=0) \sim \frac{\langle k\rangle}{\langle k^2\rangle} \sim N^{-(3-\gamma)/2}\,,
    \qquad 2<\gamma < 3\,.
\label{eq:betac_SF_g23}
\end{equation}
In this regime, we find  $\beta_c(\infty)/\beta_c(0)=\alpha/(\alpha-1)\to 1$ as $N\to\infty$,  the Hopf threshold becomes independent of the feedback strength due to network heterogeneity, and the system essentially behaves  as if $\beta_c$ were  fixed by topology alone. A representative  phase diagram for $\gamma=2.1$ is shown in the central panel of Fig.~\ref{fig:phase_diagrams}.

Finally, we analyze the case of log-normal degree distribution, where $\log k \sim \mathcal{N}(\mu,\sigma^2)$ and the moments are given by $\langle k^n \rangle
= \exp\left(n\mu+\frac{n^2\sigma^2}{2}\right)$, 
leading to 
\begin{equation}
    \alpha=\exp^{\left(\mu+\frac{3}{2}\sigma^2\right)}
\end{equation}
In Fig. \ref{fig:phase_diagrams} (right panel), we show the case $\langle k\rangle=25$ and $\sigma=1$ ($\mu = \ln\langle k\rangle-\sigma^2/2$). As in the previous case, when the presence of hubs is not negligible we have $\frac{\beta_{(c=0)}}{\beta_{(c\to \infty)}}\sim 1$.

\subsection{Kinematic waves}
\label{sec:kinematic_waves}

In networks with heterogeneous degree distributions, the response of spins to perturbations is not uniform. 
The first spins to flip are typically those associated with low-degree nodes, which have fewer neighbors, experience weaker effective fields, and are therefore more susceptible to fluctuations or external perturbations. Once these spins flip, the perturbation propagates across the network, gradually reaching nodes with higher degree. The latter are stabilized by the stronger mean field generated by their many neighbors and respond more slowly, acting as inertial elements of the dynamics. The result is a sequential, degree-ordered activation---a kinematic wave sweeping from the network periphery toward the hubs.

This  \emph{kinematic} wave  
arises from the (spatially) degree-dependent timing of a local excitation process as~\cite{Murray2002}, rather than from the transport of a physical quantity through the system. Each degree class responds independently to the same global mean field, but with a $k$-dependent delay; the apparent propagation from low to high degree is a consequence of this graded susceptibility, not of a diffusive or advective coupling between degree classes. This is analogous to kinematic waves in excitable media, where a wavefront sweeps through a spatially heterogeneous tissue because different regions reach threshold at different times, rather than because one region triggers the next.

Such kinematic waves  appear in feedback  Ising models on heterogeneous networks below $T_c$ as a degree-dependent phase offset in the oscillatory trajectories of the degree classes (Fig.~\ref{fig:panel}C), where low-degree nodes lead the cycle and high-degree nodes lag behind. The same phenomenon persists \emph{above} $T_c$, in the subcritical resonant regime where the paramagnetic fixed point is linearly stable but the leading eigenvalues of the Jacobian are complex: fluctuations are amplified into damped oscillations whose degree-resolved structure carries a measurable signature of the kinematic wave.

\begin{figure*}[t]
    \centering
    \includegraphics[width=0.4\textwidth]{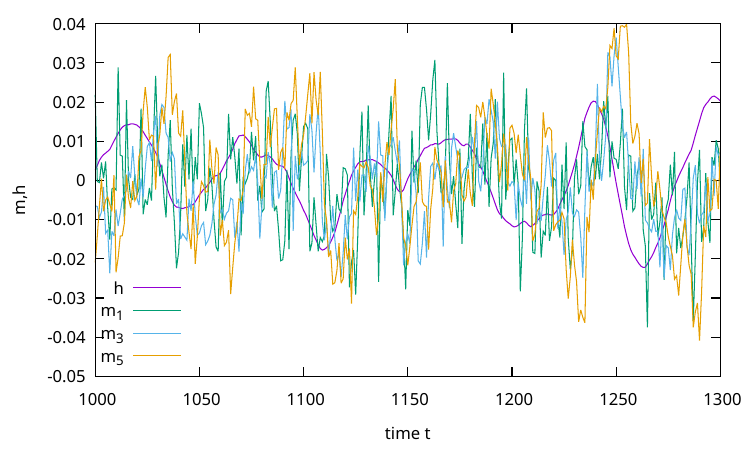}
    \includegraphics[width=0.4\textwidth]{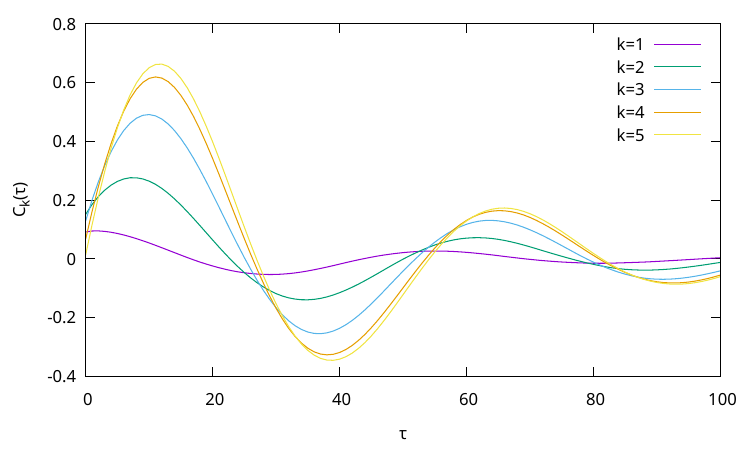}
    \includegraphics[width=0.4\textwidth]{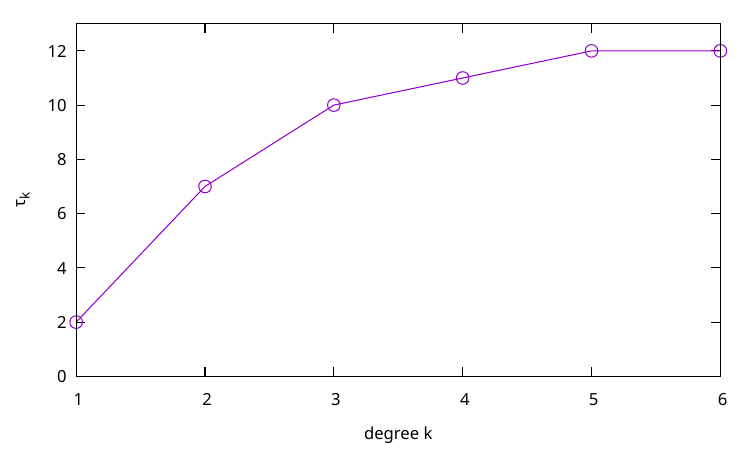}
    \caption{{\bf Kinematic wave in the subcritical resonant phase.}
    Microscopic numerical simulations on a Poissonian network with $z=4$, $c=0.1$, $\beta=0.24$, $N=10^4$.
    \textcolor{black}{\textit{Top left:}} Time series of the external field $h$ and degree-resolved magnetizations $m_1$, $m_3$, $m_5$, showing noisy oscillations with degree-dependent amplitude.
    \textcolor{black}{\textit{Top right:}} Stationary cross-correlation $C_k(\tau)=\langle m_k(t+\tau)\,h(t)\rangle$ for $k=1,\ldots,5$. The peak amplitude grows with $k$ and the peak position $\tau^*_k$ shifts to larger $\tau$ for higher degree.
    \textcolor{black}{\textit{Bottom:} } Peak delay $\tau^*_k$ as a function of degree $k$, showing the monotonic increase that characterizes the kinematic wave: low-degree nodes respond first, high-degree nodes follow with a delay.}
    \label{fig:kinematic}
\end{figure*}

Figure~\ref{fig:kinematic} illustrates this regime for a Poissonian network with $z=4$, $\beta=0.24$, $c=0.1$ and $N=10^4$ nodes. The left panel shows the time series of the external field $h(t)$ together with three representative degree-class magnetizations $m_1$, $m_3$, $m_5$. Although no macroscopic limit cycle exists (all oscillations decay on a timescale set by the damping rate of the complex eigenvalue pair), the stochastic dynamics sustains persistent noisy oscillations whose amplitude and phase depend systematically on the degree.

To quantify the waves, we measure the time-delayed cross-correlation between the feedback field and the degree-resolved magnetization, 
\begin{equation}
    C_k(\tau) = \langle m_k(t+\tau)\,h(t)\rangle\,,
    \label{eq:Ck_def}
\end{equation}
where the average is taken over the stationary stochastic process. The right top  panel of Fig.~\ref{fig:kinematic} shows $C_k(\tau)$ for $k=1,\ldots,5$. Two features stand out. First, the peak amplitude increases with $k$: high-degree nodes are more strongly correlated with the field because they couple to more neighbors and thus track the mean-field variable $u$ more faithfully. Second, the peak position $\tau^*_k$ shifts to larger lags with increasing $k$: in response to a fluctuation of $h$ at time $t$, low-degree nodes react first (peak at $\tau^*_1\approx 2$~sweeps) while high-degree nodes respond later ($\tau^*_6\approx 11$~sweeps). This is the quantitative fingerprint of the kinematic wave. In Fig.~\ref{fig:kinematic} (bottom), we show  the peak delay $\tau^*_k$ as a function of $k$. The $\tau^*_k$ increases rapidly  for small $k$ and saturates above $k\approx 4$, close to the mean degree. This saturation reflects the fact that for $k\gg z$, all high-degree nodes are already strongly locked to the mean field and respond on a similar timescale.

\subsection{Phase separation}
\label{sec:phase_sep}

\begin{figure*}[t]
    \centering
    \includegraphics[width=0.8\textwidth]{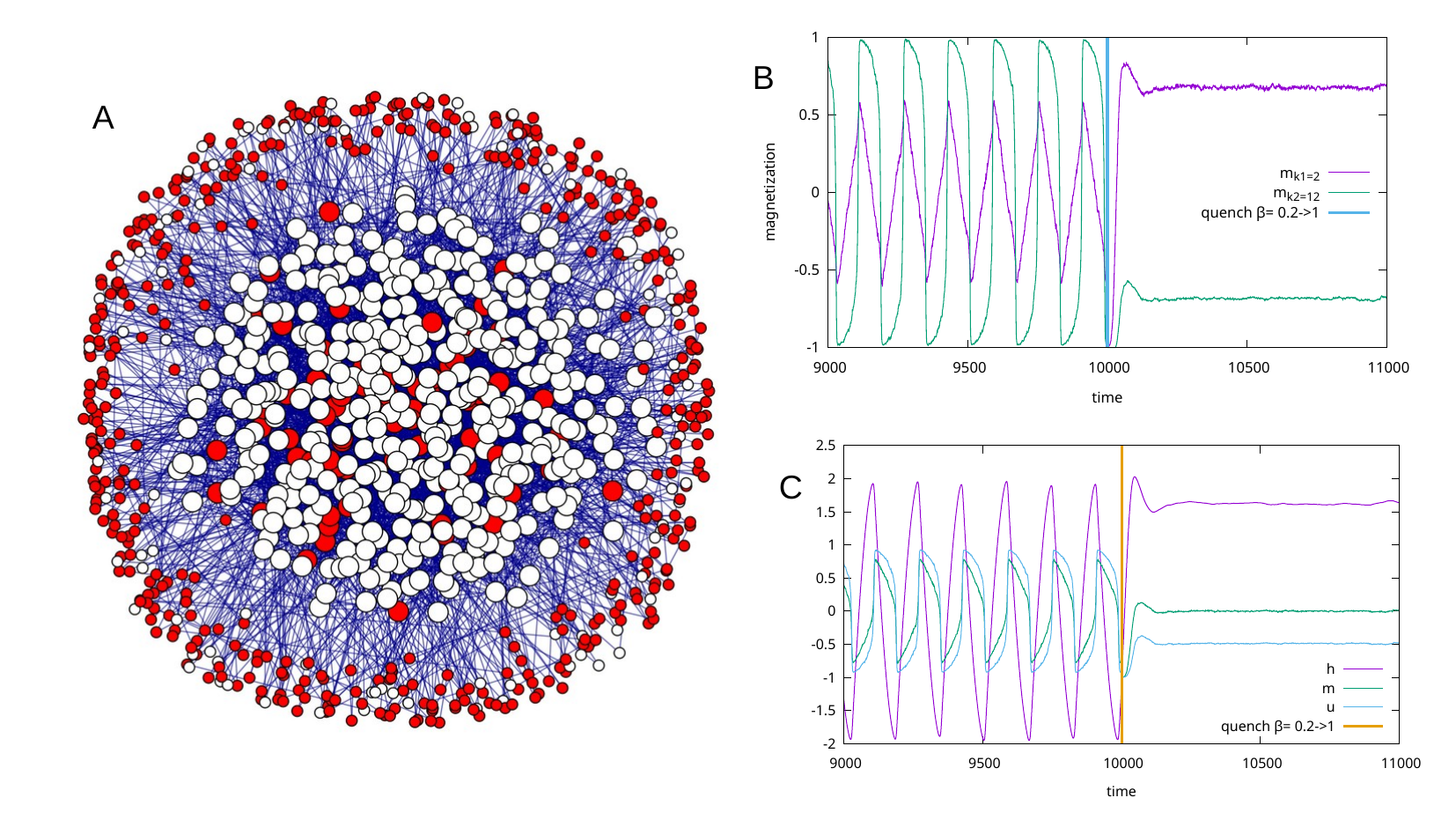}
    \caption{{\bf Phase separation after a quench.}
    Microscopic numerical simulations on a Bimodal network with $k_1=4$, $k_2=12$ $c=0.1$, $N=10^4$.
    \textit{Left:} Snapshot of the system after the quench.
    \textit{Right:} Magnetization  after and before the quench.}
    \label{fig:phase_sep}
\end{figure*}

\subsubsection{Fixed-point equations}

In the steady state, $\dot{h}=0$ implies $m=0$. Together with the self-consistency condition on~$u$, the fixed-point equations read
\begin{equation}
    \sum_k p_k \,\tanh\!\big(\beta(k\,u + h)\big) = 0\,,
        \label{eq:PS_m0}
\end{equation}
\begin{equation}
    u = \sum_k \frac{k\,p_k}{\langle k\rangle}\,
        \tanh\!\big(\beta(k\,u + h)\big)\,.
        \label{eq:PS_u}
\end{equation}
The trivial solution $u=h=0$ (the paramagnetic state) always satisfies these equations. We now ask when a nontrivial ``phase-separated'' solution with $u\neq 0$, $h\neq 0$ appears, and what its properties are.

\subsubsection{Third-order expansion and bifurcation point}

Expanding $\tanh(\beta x) = \beta x - \tfrac{1}{3}\beta^3 x^3 + O(x^5)$ with $x_k = ku+h$, Eq.~\eqref{eq:PS_m0} gives at linear order $h = -\langle k\rangle\, u \equiv -z\,u$. Writing $h = -z\,u + \delta$ and substituting back, the cubic correction yields
\begin{equation}
    \delta = \frac{\beta^2\,\mu_3}{3}\,u^3 + O(u^5)\,,
    \label{eq:h_expansion}
\end{equation}
where $\mu_n = \langle(k-z)^n\rangle$ denotes the $n$-th central moment of the degree distribution. Note that for symmetric distributions ($\mu_3=0$) the relation $h=-zu$ is exact to cubic order.

Substituting~\eqref{eq:h_expansion} into~\eqref{eq:PS_u} and expanding to third order, the self-consistency equation for $u$ becomes
\begin{equation}
    u\left[1 - \frac{\beta\,\mu_2}{z} + \frac{\beta^3\,\mu_4}{3z}\,u^2\right] = 0\,,
    \label{eq:u_cubic}
\end{equation}
where $\mu_2 = \mathrm{Var}(k)$ and $\mu_4 = \langle(k-z)^4\rangle$.
The trivial root $u=0$ factors out, and a nontrivial solution appears when $\beta\mu_2/z > 1$, i.e.\ above the critical inverse temperature
\begin{equation}
    \boxed{\beta_{c2} = \frac{\langle k \rangle}{\mathrm{Var}(k)} = \frac{z}{\mu_2}\,.}
    \label{eq:beta_c2}
\end{equation}
Since $\mu_4>0$, the bifurcation is a supercritical pitchfork. The nontrivial branch emerges continuously with amplitude
\begin{equation}
    u_s^2 = \frac{3(\beta\,\mu_2 - z)}{\beta^3\,\mu_4}\,,\qquad
    h_s = -z\,u_s + \frac{\beta^2\mu_3}{3}\,u_s^3\,.
    \label{eq:PS_amplitude}
\end{equation}
Equation \ref{eq:PS_amplitude} is valid at the third order for arbitrary~$\beta > \beta_{c2}$. A second solution with $u_s<0$ and  $h_s>0$ exists by symmetry. Physically, the phase-separated state has $m=0$ but $u\neq 0$, and  
network nodes split into two groups with opposite magnetizations, separated at a boundary degree $k^*=-h_s/u_s \approx z$.

For a Poissonian graph, $\mu_2 = z$ and  $\beta_{c2}=1$ regardless of~$z$, and  the phase separation lies deep in the low-temperature regime, far below the Hopf line $\beta_{\rm Hopf}$ given by Eq.~\eqref{eq:betac_new}. In contrast, for distributions with large variance, the two lines can approach each other: $\beta_{c2} = z/\mathrm{Var}(k) \to 0$ as $\mathrm{Var}(k)\to\infty$.

\subsubsection{Zero-temperature limit}

In the limit $\beta\to\infty$, $\tanh\to\mathrm{sgn}$ and Eq.~\eqref{eq:PS_m0} becomes
\begin{equation}
    -\!\!\sum_{k<k^*} p_k + \sum_{k>k^*} p_k = 0\,,
\end{equation}
so that $k^*$ is the \emph{median} of the degree distribution. The corresponding solution is
\begin{equation}
    u_s = \frac{1}{\langle k\rangle}\left(\sum_{k>k^*} k\,p_k - \sum_{k<k^*} k\,p_k\right),
    \qquad h_s = -k^*\,u_s\,.
\end{equation}
In this limit, the nodes orientation splits into two groups, where nodes of degree $k>k^*$ and $k<k^*$ take opposite magnetization.
\subsubsection{Stability of the phase-separated fixed point}

The Jacobian of the full dynamical system~\eqref{Eq.systEq} evaluated at the phase-separated (PS) point $(m=0,\,u_s,\,h_s)$ is
\begin{equation}
    \mathbf{J}_{\rm PS} =
    \begin{pmatrix}
        -1 & \beta\,a_1 & \beta\,a_0 \\[2pt]
        0 & -1 + \dfrac{\beta\,a_2}{z} & \dfrac{\beta\,a_1}{z} \\[6pt]
        -c & 0 & 0
    \end{pmatrix}\,,
    \label{eq:J_PS}
\end{equation}
where the \emph{effective moments} are
\begin{equation}
    a_n \equiv \sum_k p_k\,k^n\,\mathrm{sech}^2\!\big(\beta(k\,u_s+h_s)\big),
    \quad n=0,1,2\,.
    \label{eq:effective_moments}
\end{equation}
Equation~\eqref{eq:J_PS} has exactly the same structure as the Jacobian at the paramagnetic point, with $\langle k^n\rangle$ replaced by the weighted moments~$a_n$.
Its characteristic polynomial is $\lambda^3 + A\lambda^2 + B\lambda + C = 0$ with
\begin{align}
    A &= 2 - \frac{\beta\,a_2}{z}\,,\label{eq:PS_A}\\
    B &= 1 + \beta\,a_0\,c - \frac{\beta\,a_2}{z}\,,\label{eq:PS_B}\\
    C &= \frac{\beta\,c}{z}\!\left(a_0\,z - \beta(a_0\,a_2-a_1^2)\right).\label{eq:PS_C}
\end{align}
By the Routh--Hurwitz criterion, the fixed point is stable if and only if $A>0$, $C>0$, and $AB>C$; as in the paramagnetic case, it is the combined inequality $AB>C$ that sets the effective stability boundary. \textcolor{black}{When it emerges, the PS point  ($\beta=\beta_{c2}$, $u_s=0$) is unstable, since  $a_n=\langle k^n\rangle$ and $A = (\mu_2-z^2)/\mu_2 <0$ ($z^2>\mathrm{Var}(k)$ always).}

\subsubsection{Stability within the third-order expansion}

Substituting $\mathrm{sech}^2(\beta x_k) = 1-\beta^2(k-z)^2u_s^2+O(u_s^4)$ and the expression~\eqref{eq:PS_amplitude} for $u_s^2$, the effective moments become
\begin{align}
    a_0 &= 1 - \frac{3\mu_2(\beta\mu_2-z)}{\beta\,\mu_4}\,,\\
    a_1 &= z - \frac{3(\mu_3+z\mu_2)(\beta\mu_2-z)}{\beta\,\mu_4}\,,\\
    a_2 &= \langle k^2\rangle - \frac{3(\mu_4+2z\mu_3+z^2\mu_2)(\beta\mu_2-z)}{\beta\,\mu_4}\,.
\end{align}
The dominant stability condition is $A>0$, i.e.\ $\beta\,a_2/z<2$. Expanding in $\epsilon=\beta-\beta_{c2}$ we find 
\begin{multline}
    A(\epsilon) = \underbrace{\frac{\mu_2-z^2}{\mu_2}}_{A_0\,<\,0}\\
    + \frac{3\mu_2^2 z^2+6\mu_2\mu_3 z+(2\mu_2-z^2)\mu_4}{z\,\mu_4}\,\epsilon
    + O(\epsilon^2)\,.
\end{multline}
Setting $A=0$ yields the stabilization threshold
\begin{equation}
    \boxed{\epsilon_{\rm stab} = \beta_{\rm stab}-\beta_{c2}
    = \frac{(z^2-\mu_2)\,z\,\mu_4}
    {\mu_2\,\mathcal{D}}\,,}
    \label{eq:eps_stab}
\end{equation}
where $\mathcal{D} \equiv 3\mu_2^2 z^2+6\mu_2\mu_3 z+(2\mu_2-z^2)\mu_4$.
Since $z^2>\mu_2$ and the denominator is positive for all distributions we have examined, $\epsilon_{\rm stab}>0$ and the PS point always stabilizes at a temperature below its birth.
For Poisson ($\mu_2=z$, $\mu_3=z$, $\mu_4=3z^2+z$) this gives
$\epsilon_{\rm stab}=(z-1)(3z+1)/(11z+2)$,
while for a symmetric bimodal graph with $\Delta k = k_2-k_1$ and $\mu_3=0$,
$\epsilon_{\rm stab}=2z(4z^2-\Delta k^2)/[\Delta k^2(4z^2+\Delta k^2)]$.
$$
$$

\subsubsection{Bimodal random graph: exact finite-$\beta$ solution}

For a bimodal graph with degrees $k_1<k_2$, $p_1=p_2=\tfrac{1}{2}$, and $z=(k_1+k_2)/2$, the phase-separated fixed point can be solved in closed form at arbitrary~$\beta$. The $m=0$ condition~\eqref{eq:PS_m0} requires $\tanh(\beta(k_1 u+h))+\tanh(\beta(k_2 u+h))=0$, which by the odd symmetry of $\tanh$ gives exactly $h=-zu$. Substituting into~\eqref{eq:PS_u} yields the self-consistent equation
\begin{equation}
    u = \frac{\Delta k}{2z}\,\tanh\!\left(\frac{\beta\,\Delta k\,u}{2}\right),
    \label{eq:u_bimodal}
\end{equation}
with $\Delta k = k_2-k_1$. A nontrivial solution exists for $\beta>\beta_{c2}=4z/\Delta k^2 = 2(k_1+k_2)/(k_2-k_1)^2$, recovering~\eqref{eq:beta_c2} with $\mu_2=\Delta k^2/4$.

A crucial simplification arises for the stability analysis: since $h=-zu$ exactly, the local fields satisfy $\beta(k_1 u+h)=-\beta(k_2 u+h)$, so
\begin{equation}
    \mathrm{sech}^2\!\big(\beta(k_1 u_s+h_s)\big) = \mathrm{sech}^2\!\big(\beta(k_2 u_s+h_s)\big) \equiv s\,.
\end{equation}
All effective moments~\eqref{eq:effective_moments} then factor as $a_n = \langle k^n\rangle\,s$, and the Jacobian~\eqref{eq:J_PS} becomes identical to the paramagnetic Jacobian with the single substitution
\begin{equation}
    \beta \;\longrightarrow\; \phi \equiv \beta\,s = \beta\,\mathrm{sech}^2\!\!\left(\frac{\beta\,\Delta k\,u_s}{2}\right).
    \label{eq:phi_bimodal}
\end{equation}
The PS point is therefore stable whenever $\phi$ satisfies the same Routh--Hurwitz conditions that define the Hopf bifurcation of the paramagnetic state, i.e.\ when $\phi < \beta_{\rm Hopf}(c)$ with $\beta_{\rm Hopf}(c)$ given by Eq.~\eqref{eq:betac_new}. Since $\phi(\beta)$ is a monotonically decreasing function of~$\beta$ for $\beta>\beta_{c2}$ (starting at $\phi=\beta_{c2}$ and approaching zero as $\beta\to\infty$), the stabilization condition reduces to
\begin{equation}
    \phi(\beta_{\rm stab}) = \beta_{\rm Hopf}(c)\,.
    \label{eq:stab_bimodal}
\end{equation}

\section{Conclusion}

We have studied the feedback Ising model on random graphs with heterogeneous degree distributions, developing a Curie--Weiss (heterogeneous mean-field) theory that yields analytical phase diagrams in good agreement with Monte Carlo simulations. The equilibrium Ising model on such graphs had been rigorously characterized through heterogeneous mean-field and replica methods to show how  the degree distribution controls the onset of ferromagnetic order~\cite{leone2002ferromagnetic,Dorogovtsev2002,Goltsev2003}. In the present work, we have investigated how the feedback modifies this picture.


We found that the equilibrium ferromagnetic transition is replaced by an Andronov--Hopf bifurcation at a critical inverse temperature $\beta_c(c)$ given in closed form by Eq.~\eqref{eq:betac_new}. This expression depends on the degree distribution only through the moment ratio $\alpha = \langle k^2\rangle/\langle k\rangle$ and interpolates between the classical mean-field critical point $\beta_c = 1/\alpha$ for vanishing feedback and a saturation value $\beta_c = 1/(\alpha-1)$ for strong feedback. On scale-free networks with diverging second moment ($2<\gamma\leq 3$), the system is in the oscillatory phase at any nonzero temperature in the thermodynamic limit, and the Hopf threshold becomes insensitive to the feedback strength. 

Degree heterogeneity promotes the emergence of two distinct order parameters: the spiking rate per neuron $m$ and the spiking rate per synapse $u$, which coincide on regular graphs but decouple on heterogeneous networks. This echoes the central lesson  of statistical mechanics on heterogeneous networks, namely  that the relevant order parameter is not the naive average, but  a degree-weighted quantity~\cite{leone2002ferromagnetic,
Dorogovtsev2002,Goltsev2003}. 
The interplay between $m$ and $u$ gives rise to two phenomena that have no counterpart in the homogeneous case: kinematic waves and phase-separated fixed points. 

Kinematic waves are degree-ordered sequential activations in which low-degree nodes respond before high-degree hubs. We found them both in the limit-cycle phase (as degree-dependent phase offsets) and in the subcritical resonant regime, as $k$-dependent peak delays in the cross-correlation $\langle m_k(t+\tau)\,h(t)\rangle$. The wave structure is encoded in the Hopf eigenvector through the single complex ratio $q_u/q_h = \beta_H/(\alpha+i\omega_0)$, which determines both the amplitude hierarchy and the phase ordering across degree classes. We stress that these are kinematic waves in the sense of Murray~\cite{Murray2002}: they arise from graded susceptibility across degree classes, not from transport of a physical quantity.

 The phase-separated fixed point with $m=0$ but $u\neq 0$ emerges via a supercritical pitchfork bifurcation at $\beta_{c2} = \langle k\rangle/\mathrm{Var}(k)$. In this state, degree classes split into two groups with opposite magnetizations separated at a boundary degree $k^*$ near the median of the distribution. The fixed point is always born as an unstable saddle but can stabilize at lower temperatures. For bimodal graphs, the stability analysis simplifies dramatically: the Jacobian at the phase-separated point is identical to the paramagnetic one with the replacement $\beta \to \phi = \beta\,\mathrm{sech}^2(\beta\Delta k\,u_s/2)$, yielding the exact stabilization condition $\phi(\beta_{\rm stab}) = \beta_{\rm Hopf}(c)$. When the phase-separated point stabilizes, it captures the dynamics and destroys the limit cycle. For Poissonian graphs, the phase-separated point remains a saddle at all temperatures within the third-order expansion, and the limit cycle persists.

Several directions remain open for future work. A degree-resolved spectral analysis reveals that the $(K+1)$-dimensional linearized system reduces exactly to the same cubic characteristic equation as the three-variable $(m,u,h)$ projection, plus $K-2$ degenerate ``dark modes'' at eigenvalue $-1$. This degeneracy is a consequence of the annealed (Curie--Weiss) approximation, in which the interaction matrix is rank one. On quenched graphs, this degeneracy would be lifted and the dark modes would acquire nontrivial relaxation rates, potentially giving rise to richer kinematic-wave structures. A systematic treatment beyond the naive dynamical Curie--Weiss mean field, for instance through cavity methods \cite{aurell2017cavity,dominguez2020cavity,aurell2023closure,ortega2022dynamics}, or direct dynamical mean field theories / path integral methods \cite{metz2025dynamical, aguilera2021unifying,aguilera2023nonequilibrium,aguilera2025explosive} would be valuable to assess the robustness of the phenomena reported here.

We emphasize that our analysis has focused on the simplest oscillatory instability, the Andronov--Hopf bifurcation and the resulting limit cycles, as the primary dynamical scenario.
Detailed studies of low-dimensional neuron models such as the Hindmarsh--Rose system and/or phenomenological homogeneous mean field have revealed a far richer atlas of bifurcations, including homoclinic and heteroclinic connections, spike-adding cascades, fold/hom and fold/Hopf bursting transitions, and routes to chaos through period-doubling and Shilnikov-type mechanisms~\cite{Barrio2020homoclinic,Barrio2020spikeadding,Barrio2021classification,Serrano2021order,Barrio2024geometry}.
These complex dynamical scenarios arise naturally in deterministic low-dimensional systems and can produce qualitatively different collective behaviors, such as bursting, mixed-mode oscillations, and chaotic spiking, that go well beyond the simple limit-cycle picture considered here.
However, the strength of the present approach lies precisely in the complementary direction: by working within the framework of statistical mechanics on random graphs, we can systematically address the effects of structural heterogeneity (arbitrary degree distributions), quenched disorder, and finite-size stochastic fluctuations, which are typically absent from the dynamical-systems analyses of individual neuron models or small deterministic circuits.
We believe that an important avenue for future work is to integrate these two perspectives: on the one hand, extending the   framework to incorporate richer local dynamics (e.g. beyond  binary spins) while retaining the analytical tractability afforded by the heterogeneous mean-field approach; on the other hand, studying how the complex bifurcation structures identified in neuron models are modified by network heterogeneity and noise.

A more richer dynamical bifurcation scenario would also increase the possibility of modeling collective behaviors emerging in biological neural networks. For instance,   tricritical Ising and 3-state feedback Potts model \cite{sinelshchikov2023emergence}
could show noise-induced switching between the limit cycle and the paramagnetic or phase-separated states in the coexistence regime near the Bautin point. Verifying this prediction in simulations would connect the bifurcation analysis to the statistics of intermittent oscillatory episodes, which are ubiquitous in neural recordings. Empirical work on cortical recordings has revealed that neuronal  avalanches and oscillations can coexist near a non-equilibrium  critical point~\cite{lombardi2023statistical}, and systematic analysis of neuronal population data suggests that the brain  operates near a genuine out-of-equilibrium phase transition  separating asynchronous and synchronous cortical states~\cite{Fontenele2019,Copelli2020}.
Self-organization toward criticality through adaptive and plastic synaptic mechanisms~\cite{Levina2007,Levina2009,Safavi2024,fl2012,fl2017chaos,lda:06} provides a dynamical  underpinning for why neural networks may operate near such  transitions, and connects naturally to the homeostatic feedback  loop studied here, which can also mimic network  inhibitory effects.
These findings indicate that biological neural circuits may exploit  multiple types of criticality simultaneously equilibrium-like  scaling in avalanche statistics, out-of-equilibrium Hopf-type  synchronization, and potentially more exotic non-equilibrium 
universality classes. Understanding which type of criticality is relevant in a given  context, and how network heterogeneity modifies the associated universality class, remains an important open question that connects  the present work to a broader statistical physics of neural dynamics.
Closely related to the present framework are recent works  on more general non-equilibrium mean-field spin models 
\cite{guislain2023nonequilibrium,
guislain2024discontinuous,guislain2024tailoring,
guislain2024collective,guislain2024hidden,
guislain2025farfromequilibrium}. These studies reveal that even in homogeneous mean-field settings,  the interplay between non-reciprocal interactions and stochastic  dynamics can produce continuous or discontinuous transitions to  oscillatory phases, hidden collective oscillations in disordered  models, and complex energy landscapes far from equilibrium. Extending  oscillating spin models to heterogeneous graphs  constitutes a natural and promising direction for future work. Finally, the observation that the Hopf threshold depends on the degree distribution through $\alpha = \langle k^2\rangle/\langle k\rangle$ raises the question of whether a slow drift of this quantity (e.g., through gradual loss of nodes or synapses, as occurs in brain aging) could drive the system across the phase diagram. The framework developed here provides analytical tools to address this question quantitatively.
\begin{acknowledgments}
D.D.M.\ acknowledges financial support from the grants
PIBA\_2024\_1\_0016 (Basque Government) and Project
PID2023-146408NB-I00 funded by MICIU/AEI/10.13039/501100011033
and by FEDER, UE.
\end{acknowledgments}
\clearpage
\onecolumngrid
\appendix

\section
{Van Kampen, system-size expansion}
\label{App.VanKampen}
In this section, we provide the derivation of the dynamical equations for $m_k$ and $u$ proposed in the paper. We then start from the definition of the master equation in the mean field limit to finally write down the Fokker-Planck and the Langevin equations. Let us describe with $\,\boldsymbol{s}=(s_1,\dots,s_N)$ a given configuration of the spins, where $s_i=\pm1$, and $P(\boldsymbol{s},t)$ the probability of $\boldsymbol{s}$. We then suppose the system to follow a Glauber dynamics with spin flip rates
\begin{equation}
  w_i(\boldsymbol{s})=\frac{1}{2}\Big[1 - s_i \tanh\big(\beta\, H_i^{\rm eff}(\boldsymbol{s})\big)\Big].
\end{equation}
Here $H_i^{\rm eff}$ is the local field on site $i$. For an Ising model on an undirected and unweighted graph $H_i^{\rm eff}$ reads as
\begin{equation}
  H_i^{\rm eff}(\boldsymbol{s}) = \sum_{j} A_{ij}\, s_j + h,
\end{equation}
where $A_{ij}$ is the  adjacency matrix. We suppose the coupling to be homogeneous over the graph and equal to 1.
At first, we can write the master equation as
\begin{equation}
  \frac{d}{dt}P(\boldsymbol{s},t)=\sum_{i=1}^N 
  \Big[ w_i(\boldsymbol{s}^i)\,P(\boldsymbol{s}^i,t) - w_i(\boldsymbol{s})\,P(\boldsymbol{s},t)\Big],
\end{equation}
with $\boldsymbol{s}^i$ standing for the configuration where the spin $i$ is flipped.
\\Here we introduce the mean-field approximation, as carried out in \cite{leone2002ferromagnetic}. Then let us introduce the average magnetization per degree $k$
\begin{equation}
  m_k(t)=\frac{1}{N_k}\sum_{i\in\mathcal{V}_k} s_i(t)
\end{equation}
and the average magnetization seen on a nearest neighbor node, given by
\begin{equation}
     u(t) \equiv \frac{\sum_k k\,p_k\, m_k(t)}{\langle k\rangle}.
\end{equation}
In this light, the effective Hamiltonian takes the form $H_i^{\rm eff}\approx k\,u(t)+h(t)$ and the average flip rate for the sites with degree $k$ is
\begin{align}
  W_+^{(k)} &= \frac{N_k}{2}\,(1-m_k)\,\Big[1+\tanh\big(\beta(k u+h)\big)\Big],\\
  W_-^{(k)} &= \frac{N_k}{2}\,(1+m_k)\,\Big[1-\tanh\big(\beta(k u+h)\big)\Big],
\end{align}
where $\frac{N_k}{2}\,(1-m_k)$ and $\frac{N_k}{2}\,(1+m_k)$ are respectively the fraction of degree
$k$ spins in the down and up states.
\\From here
\begin{equation}
\begin{aligned}
\frac{\partial}{\partial t} P(\{m_k\},t) \;=\; 
&\sum_{k} \Big\{ 
    W_+^{(k)}(m_k-\Delta m_k,\,u(\dots,m_k-\Delta m_k,\dots))\; 
      P(\dots,m_k-\Delta m_k,\dots;\,t) \\
&\qquad\qquad +\; W_-^{(k)}(m_k+\Delta m_k,\,u(\dots,m_k+\Delta m_k,\dots))\; 
      P(\dots,m_k+\Delta m_k,\dots;\,t) \\
&\qquad\qquad -\; \big[W_+^{(k)}(m_k,u(\{m_{k'}\})) + W_-^{(k)}(m_k,u(\{m_{k'}\}))\big]\;
      P(\{m_k\},t)
\Big\}.
\end{aligned}
\label{eq:ME}
\end{equation}
In a Kramers$-$Moyal expansion, we can derive the coefficients of the Fokker-Planck equation through Eq. \ref{eq:ME} as $E(\Delta m_k|\{m_{k^{\prime}}\})/\Delta t$ and $E(\Delta m^2_k|\{m_{k^{\prime}}\})/\Delta t$. If we keep that each flip implies $\Delta m_k=\pm 2/N_k$, we obtain $E(\Delta m_k|\{m_{k^{\prime}}\})/\Delta t = \frac{2}{N_k}\big(W_+^{(k)}-W_-^{(k)}\big) $ and $E(\Delta m^2_k|\{m_{k^{\prime}}\})/\Delta t =  \frac{4}{N_k^2}\big(W_+^{(k)}+W_-^{(k)}\big)$. Given that the fluctuation scales with $1/N_k$ with respect to the average rate, we assume here to apply a Van Kampen expansion. The multivariate Fokker-Planck equation reads as
\begin{equation}
  \partial_t P 
  = -\sum_i \partial_{x_i}\!\left[\frac{2}{N_k}\big(W_+^{(k)}-W_-^{(k)}\big)\,P\right]
    + \frac{1}{2}\sum_{i,j}\partial_{x_i}\partial_{x_j}\!\left[B_{ij}(\mathbf{x})\,P\right],
\end{equation}
where $B_{ij}(\mathbf{x})$ is the diffusion matrix. We assume the spin flip to be independent among different degrees $k\neq k'$, then
\begin{equation}
  B_{m_k m_{k'}}(\mathbf{x})=0 \qquad (k\neq k').
\end{equation}
All considered we obtain 
\begin{equation}
  \partial_t P 
  = -\sum_k \partial_{k}\!\left[-\,m_k + \tanh\!\big(\beta(k u+h)\big)\,P\right]
    + \frac{1}{2}\sum_{k, k^{\prime}}\partial_{k}\partial_{k^{\prime}}\!\left[\frac{2}{N_k}\,\Big[1 - m_k \tanh\!\big(\beta(k u+h)\big)\Big]\,P\right].
\end{equation}
Finally we can derive the corresponding Langevin equations. 
\begin{equation}
    \dot m_k(t) = -\,m_k(t) + \tanh\!\big(\beta(k u(t) + h(t))\big)
              \;+\; \sqrt{\frac{2}{N_k}\,\Big[1 - m_k \tanh\!\big(\beta(k u+h)\big)\Big]}\;\xi_k(t),
\qquad k=1,\dots,K, \label{eq:L_mk} 
\end{equation}
where \(\{\xi_k(t)\}_{k=1}^K\) stands for the withe noise variables
\[
\langle \xi_k(t) \rangle = 0,\qquad 
\langle \xi_k(t)\,\xi_{k'}(t')\rangle = \delta_{k k'}\,\delta(t-t')\].
In conclusion, the deterministic dynamical equations can be obtained by neglecting the last term. The we end up with the following set of equations
\begin{eqnarray}
    \dot m_k &=& -m_k + \tanh\big(\beta k u + \beta h\big)\\
    \dot{h} &=& -cm
\end{eqnarray}
or alternatively
\begin{eqnarray}
\dot{m} &=& -m+\beta(\langle k \rangle u+h) \\  
\dot{u} &=& -u+\beta(\frac{\langle k^2 \rangle}{\langle k \rangle} u+h) \\
\dot{h} &=& -cm
\end{eqnarray}

\section{Phase detection pseudo-code}
\begin{figure*}[h]
    \centering
    \includegraphics[width=\textwidth]{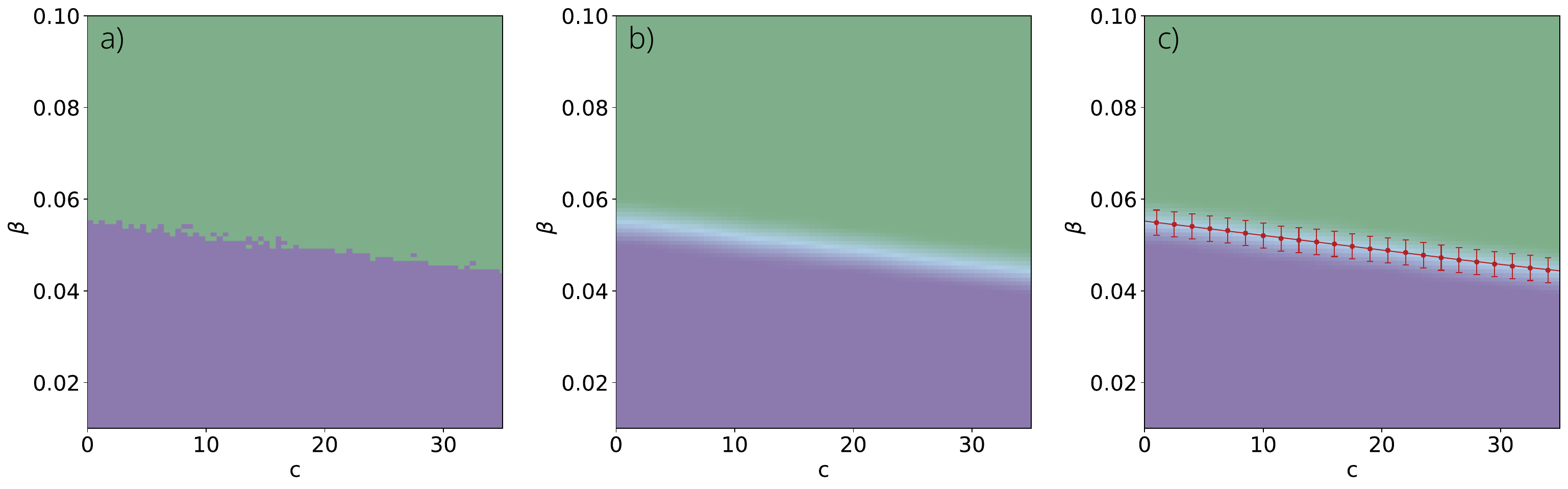}
    \caption{{\bf Computational derivation of the phase diagrams in the $(c,\beta)$ plane for the Erd\H{o}s--R\'enyi graph.} \textit{Left:} The purple and the green areas respectively correspond to the $(m=0, h=0)$ fixed point and the limit cycle detection according to the pseudo-code above. \textit{Center:} To panel a) is applied a squared gaussian filter convolution, with edge length $=11$ and $\sigma = 2$. \textit{Right:} To the final diagram is applied a polynomial fit over the points with values $1/2$ to extract the boundaries. The fit is reported with the red solid line. Finally, the cumulative error bars are reported.}
    \label{fig:GaussianFilter}
\end{figure*}

Here we provide a pseudo-code for the limit cycles detection. We consider the classification of three scenarios: the existence of a limit cycle, of a fixed point in $(m=0, h=0)$ and a fixed point in $(m=0, h\not =0)$. To distinguish the three we first analyze the signal through a Hilbert transformation, this allows us to detect the presence of a limit cycle. If the limit cycle is absent we evaluate the kind of fixed point by looking at the time average of the magnetic field $\bar{h}$. A pseudo-code is provided in the following, where phase$=\{0,1,2\}$ respectively stand for the fixed point in $(m=0, h=0)$, the limit cycle and the fixed point in $(m=0, h\not =0)$.\\\\
\begin{center}
\begin{tcolorbox}[width=0.7\textwidth]
\noindent
\vspace*{0.1cm}for each file ($\beta,c$):
\\\vspace*{0.1cm}\hspace*{0.5cm}compute analytic signal via Hilbert transformation
\\\vspace*{0.1cm}\hspace*{0.5cm}A = instantaneous amplitude
\\\vspace*{0.1cm}\hspace*{0.5cm}estimate p(A) using KDE
\\\vspace*{0.1cm}\hspace*{0.5cm}find local maxima of p(A)
\\\vspace*{0.1cm}\hspace*{0.5cm}max peak position = max$\{A:\, A=\text{ point of local maximum of }p(A)\}$
\\\vspace*{0.1cm}\hspace*{0.5cm}if no peak:
\\\vspace*{0.1cm}\hspace*{1cm}phase = nan
\\\vspace*{0.1cm}\hspace*{0.5cm}else if max peak position $> 0.05$ and $p(0) < 1$:
\\\vspace*{0.1cm}\hspace*{1cm}phase = 1
\\\vspace*{0.1cm}\hspace*{0.5cm}else if mean(h) $< 1$:
\\\vspace*{0.1cm}\hspace*{1cm}phase $= 0$
\\\vspace*{0.1cm}\hspace*{0.5cm}else:
\\\vspace*{0.1cm}\hspace*{1cm}phase $= 2$
\end{tcolorbox}
\end{center}
The peak\_threshold and the $p(0)$\_threshold are experimentally derived from the signals observation and are respectively peak\_threshold$=0.05$ and $p(0)$\_threshold$=1$.
\\With this procedure we derive a map as in Fig. \ref{fig:GaussianFilter} panel a). We then apply a gaussian filter as done done in panel b). Finally, the bifurcation line is obtained fitting the points where the map takes the value $1/2$, reporting the error bars (see panel c)).

\def\url#1{}
\bibliography{cite}

@article{sinelshchikov2023emergence,
  title={Emergence of collective self-oscillations in minimal lattice models with feedback},
  author={Sinelshchikov, Dmitry and Poggialini, Anna and Abbate, Maria Francesca and De Martino, Daniele},
  journal={Physical Review E},
  volume={108},
  number={4},
  pages={044204},
  year={2023},
  publisher={APS}
}

@article{Bullmore2009,
  author  = {Bullmore, Ed and Sporns, Olaf},
  title   = {Complex brain networks: graph theoretical analysis of structural and functional systems},
  journal = {Nature Reviews Neuroscience},
  volume  = {10},
  pages   = {186--198},
  year    = {2009},
  doi     = {10.1038/nrn2575}
}

@book{Sporns2011,
  author    = {Sporns, Olaf},
  title     = {Networks of the Brain},
  publisher = {MIT Press},
  year      = {2011}
}

@article{Hopfield1982,
  author  = {Hopfield, J. J.},
  title   = {Neural networks and physical systems with emergent collective computational abilities},
  journal = {Proceedings of the National Academy of Sciences},
  volume  = {79},
  pages   = {2554--2558},
  year    = {1982},
  doi     = {10.1073/pnas.79.8.2554}
}

@article{Amit1985,
  author  = {Amit, Daniel J. and Gutfreund, Hanoch and Sompolinsky, Haim},
  title   = {Spin-glass models of neural networks},
  journal = {Physical Review A},
  volume  = {32},
  pages   = {1007--1018},
  year    = {1985},
  doi     = {10.1103/PhysRevA.32.1007}
}

@book{Buzsaki2006,
  author    = {Buzs{\'a}ki, Gy{\"o}rgy},
  title     = {Rhythms of the Brain},
  publisher = {Oxford University Press},
  year      = {2006}
}

@article{vanVreeswijk1996,
  author  = {van Vreeswijk, Carl and Sompolinsky, Haim},
  title   = {Chaos in neuronal networks with balanced excitatory and inhibitory activity},
  journal = {Science},
  volume  = {274},
  pages   = {1724--1726},
  year    = {1996},
  doi     = {10.1126/science.274.5293.1724}
}

@article{Brunel2000,
  author  = {Brunel, Nicolas},
  title   = {Dynamics of sparsely connected networks of excitatory and inhibitory spiking neurons},
  journal = {Journal of Computational Neuroscience},
  volume  = {8},
  pages   = {183--208},
  year    = {2000},
  doi     = {10.1023/A:1008925309027}
}

@book{Kuramoto1984,
  author    = {Kuramoto, Yoshiki},
  title     = {Chemical Oscillations, Waves, and Turbulence},
  publisher = {Springer},
  year      = {1984},
  doi       = {10.1007/978-3-642-69689-3}
}

@article{Acebron2005,
  author  = {Acebr{\'o}n, Juan A. and Bonilla, Luis L. and P{\'e}rez Vicente, Conrad J. and Ritort, F{\'e}lix and Spigler, Renato},
  title   = {The {Kuramoto} model: A simple paradigm for synchronization phenomena},
  journal = {Reviews of Modern Physics},
  volume  = {77},
  pages   = {137--185},
  year    = {2005},
  doi     = {10.1103/RevModPhys.77.137}
}

@article{Albert2002,
  author  = {Albert, R{\'e}ka and Barab{\'a}si, Albert-L{\'a}szl{\'o}},
  title   = {Statistical mechanics of complex networks},
  journal = {Reviews of Modern Physics},
  volume  = {74},
  pages   = {47--97},
  year    = {2002},
  doi     = {10.1103/RevModPhys.74.47}
}

@article{Newman2003,
  author  = {Newman, M. E. J.},
  title   = {The structure and function of complex networks},
  journal = {SIAM Review},
  volume  = {45},
  pages   = {167--256},
  year    = {2003},
  doi     = {10.1137/S003614450342480}
}

@article{Dorogovtsev2002,
  author  = {Dorogovtsev, S. N. and Goltsev, A. V. and Mendes, J. F. F.},
  title   = {Ising model on networks with an arbitrary distribution of connections},
  journal = {Physical Review E},
  volume  = {66},
  pages   = {016104},
  year    = {2002},
  doi     = {10.1103/PhysRevE.66.016104}
}

@article{Goltsev2003,
  author  = {Goltsev, A. V. and Dorogovtsev, S. N. and Mendes, J. F. F.},
  title   = {Critical phenomena in networks},
  journal = {Physical Review E},
  volume  = {67},
  pages   = {026123},
  year    = {2003},
  doi     = {10.1103/PhysRevE.67.026123}
}

@article{de2019feedback,
  title={Feedback-induced self-oscillations in large interacting systems subjected to phase transitions},
  author={De Martino, Daniele},
  journal={Journal of Physics A: Mathematical and Theoretical},
  volume={52},
  number={4},
  pages={045002},
  year={2019},
  publisher={IOP Publishing}
}

@article{de2019oscillations,
  title={Oscillations in feedback-driven systems: Thermodynamics and noise},
  author={De Martino, Daniele and Barato, Andre C},
  journal={Physical Review E},
  volume={100},
  number={6},
  pages={062123},
  year={2019},
  publisher={APS}
}

@article{aguilera2021unifying,
  title={A unifying framework for mean-field theories of asymmetric kinetic Ising systems},
  author={Aguilera, Miguel and Moosavi, S Amin and Shimazaki, Hideaki},
  journal={Nature communications},
  volume={12},
  number={1},
  pages={1197},
  year={2021},
  publisher={Nature Publishing Group UK London}
}

@article{aguilera2023nonequilibrium,
  title={Nonequilibrium thermodynamics of the asymmetric Sherrington-Kirkpatrick model},
  author={Aguilera, Miguel and Igarashi, Masanao and Shimazaki, Hideaki},
  journal={Nature Communications},
  volume={14},
  number={1},
  pages={3685},
  year={2023},
  publisher={Nature Publishing Group UK London}
}

@article{Barrio2020homoclinic,
  author  = {Barrio, Roberto and Ib{\'a}{\~n}ez, Santiago and P{\'e}rez, Luc{\'i}a},
  title   = {Homoclinic organization in the {Hindmarsh--Rose} model: A three parameter study},
  journal = {Chaos},
  volume  = {30},
  pages   = {053132},
  year    = {2020},
  doi     = {10.1063/1.5138919}
}

@article{Barrio2020spikeadding,
  author  = {Barrio, Roberto and Ib{\'a}{\~n}ez, Santiago and P{\'e}rez, Luc{\'i}a and Serrano, Sergio},
  title   = {Spike-adding structure in fold/hom bursters},
  journal = {Communications in Nonlinear Science and Numerical Simulation},
  volume  = {83},
  pages   = {105100},
  year    = {2020},
  doi     = {10.1016/j.cnsns.2019.105100}
}

@article{Barrio2021classification,
  author  = {Barrio, Roberto and Ib{\'a}{\~n}ez, Santiago and P{\'e}rez, Luc{\'i}a and Serrano, Sergio},
  title   = {Classification of fold/hom and fold/{Hopf} spike-adding phenomena},
  journal = {Chaos},
  volume  = {31},
  pages   = {043120},
  year    = {2021},
  doi     = {10.1063/5.0037942}
}

@article{Serrano2021order,
  author  = {Serrano, Sergio and Mart{\'i}nez, M. Angeles and Barrio, Roberto},
  title   = {Order in chaos: Structure of chaotic invariant sets of square-wave neuron models},
  journal = {Chaos},
  volume  = {31},
  pages   = {043108},
  year    = {2021},
  doi     = {10.1063/5.0043302}
}

@article{Fontenele2019,
  author  = {Fontenele, Antonio J. and de Vasconcelos, Nivaldo A. P.
             and Feliciano, Thaís and Aguiar, Leandro A. A.
             and Soares-Cunha, Carina and Coimbra, Bárbara
             and Dalla Porta, Leonardo and Ribeiro, Sidarta
             and Rodrigues, Ana João and Sousa, Nuno
             and Carelli, Pedro V. and Copelli, Mauro},
  title   = {Criticality between cortical states},
  journal = {Physical Review Letters},
  volume  = {122},
  pages   = {208101},
  year    = {2019},
  doi     = {10.1103/PhysRevLett.122.208101}
}

@article{Levina2007,
  author  = {Levina, Anna and Herrmann, J. Michael and Geisel, Theo},
  title   = {Dynamical synapses causing self-organized criticality 
             in neural networks},
  journal = {Nature Physics},
  volume  = {3},
  pages   = {857--860},
  year    = {2007},
  doi     = {10.1038/nphys758}
}

@article{Levina2009,
  author  = {Levina, Anna and Herrmann, J. Michael and Geisel, Theo},
  title   = {Phase transitions towards criticality in a neural system 
             with adaptive interactions},
  journal = {Physical Review Letters},
  volume  = {102},
  pages   = {118110},
  year    = {2009},
  doi     = {10.1103/PhysRevLett.102.118110}
}

@article{Safavi2024,
  author  = {Safavi, Shervin and Chalk, Matthew and 
             Logothetis, Nikos K. and Levina, Anna},
  title   = {Signatures of criticality in efficient coding networks},
  journal = {Proceedings of the National Academy of Sciences},
  volume  = {121},
  number  = {41},
  pages   = {e2302730121},
  year    = {2024},
  doi     = {10.1073/pnas.2302730121}
}

@article{Copelli2020,
  author  = {Copelli, Mauro},
  title   = {Physics of psychophysics: {S}pikes matter for 
             network-scaled {O}rnstein--{U}hlenbeck models},
  journal = {Physical Review Research},
  volume  = {2},
  pages   = {013058},
  year    = {2020},
  doi     = {10.1103/PhysRevResearch.2.013058}
}

@article{guislain2023nonequilibrium,
  author  = {Guislain, Laura and Bertin, Eric},
  title   = {Nonequilibrium phase transition to temporal oscillations in mean-field spin models},
  journal = {Physical Review Letters},
  year    = {2023},
  volume  = {130},
  number  = {20},
  pages   = {207102},
  doi     = {10.1103/PhysRevLett.130.207102}
}

@article{guislain2024discontinuous,
  author  = {Guislain, Laura and Bertin, Eric},
  title   = {Discontinuous phase transition from ferromagnetic to oscillating states in a nonequilibrium mean-field spin model},
  journal = {Physical Review E},
  year    = {2024},
  volume  = {109},
  number  = {3},
  pages   = {034131},
  doi     = {10.1103/PhysRevE.109.034131}
}

@article{guislain2024tailoring,
  author  = {Guislain, Laura and Bertin, Eric},
  title   = {Tailoring the overlap distribution in driven mean-field spin models},
  journal = {Physical Review B},
  year    = {2024},
  volume  = {109},
  number  = {18},
  pages   = {184203},
  doi     = {10.1103/PhysRevB.109.184203}
}

@article{guislain2024collective,
  author  = {Guislain, Laura and Bertin, Eric},
  title   = {Collective oscillations in a three-dimensional spin model with non-reciprocal interactions},
  journal = {Journal of Statistical Mechanics: Theory and Experiment},
  year    = {2024},
  volume  = {2024},
  number  = {9},
  pages   = {093210},
  doi     = {10.1088/1742-5468/ad72dc}
}

@article{guislain2024hidden,
  author  = {Guislain, Laura and Bertin, Eric},
  title   = {Hidden collective oscillations in a disordered mean-field spin model with non-reciprocal interactions},
  journal = {Journal of Physics A: Mathematical and Theoretical},
  year    = {2024},
  volume  = {57},
  number  = {37},
  pages   = {375001},
  doi     = {10.1088/1751-8121/ad6ab4}
}

@article{guislain2025farfromequilibrium,
  author  = {Guislain, Laura and Bertin, Eric},
  title   = {Far-from-equilibrium complex landscapes},
  journal = {Physical Review E},
  year    = {2025},
  volume  = {111},
  number  = {6},
  pages   = {L062101},
  doi     = {10.1103/PhysRevE.111.L062101}
}

@article{Barrio2024geometry,
  author  = {Barrio, Roberto and Ib{\'a}{\~n}ez, Santiago and P{\'e}rez, Luc{\'i}a},
  title   = {Exploring the geometry of the bifurcation sets in parameter space},
  journal = {Scientific Reports},
  volume  = {14},
  pages   = {10900},
  year    = {2024},
  doi     = {10.1038/s41598-024-61574-6}
}

@article{aguilera2025explosive,
  title={Explosive neural networks via higher-order interactions in curved statistical manifolds},
  author={Aguilera, Miguel and Morales, Pablo A and Rosas, Fernando E and Shimazaki, Hideaki},
  journal={Nature Communications},
  volume={16},
  number={1},
  pages={6511},
  year={2025},
  publisher={Nature Publishing Group UK London}
}

@article{dominguez2020cavity,
  title={The cavity master equation: average and fixed point of the ferromagnetic model in random graphs},
  author={Dom{\'\i}nguez, E and Machado, D and Mulet, R},
  journal={Journal of Statistical Mechanics: Theory and Experiment},
  volume={2020},
  number={7},
  pages={073304},
  year={2020},
  publisher={IOP Publishing and SISSA}
}

@article{aurell2023closure,
  title={A closure for the master equation starting from the dynamic cavity method},
  author={Aurell, Erik and Machado Perez, David and Mulet, Roberto},
  journal={Journal of Physics A: Mathematical and Theoretical},
  volume={56},
  number={17},
  pages={17LT02},
  year={2023},
  publisher={IOP Publishing}
}

@article{ortega2022dynamics,
  title={Dynamics of epidemics from cavity master equations: Susceptible-infectious-susceptible models},
  author={Ortega, Ernesto and Machado, David and Lage-Castellanos, Alejandro},
  journal={Physical Review E},
  volume={105},
  number={2},
  pages={024308},
  year={2022},
  publisher={APS}
}

@article{metz2025dynamical,
  title={Dynamical mean-field theory of complex systems on sparse directed networks},
  author={Metz, Fernando L},
  journal={Physical Review Letters},
  volume={134},
  number={3},
  pages={037401},
  year={2025},
  publisher={APS}
}

@article{aurell2017cavity,
  title={Cavity master equation for the continuous time dynamics of discrete-spin models},
  author={Aurell, Erik and Del Ferraro, Gino and Dom{\'\i}nguez, Eduardo and Mulet, Roberto},
  journal={Physical Review E},
  volume={95},
  number={5},
  pages={052119},
  year={2017},
  publisher={APS}
}

@article{Dahmen2025,
  author  = {Dahmen, David and Hutt, Axel and Indiveri, Giacomo and Kennedy, Ann and Lefebvre, Jeremie and Mazzucato, Luca and Motter, Adilson E. and Narayanan, Rishikesh and Payvand, Melika and Planert, Henrike and Gast, Richard},
  title   = {How heterogeneity shapes dynamics and computation in the brain},
  journal = {Neuron},
  year    = {2025},
  doi     = {10.1016/j.neuron.2025.11.023},
  note    = {Online ahead of print}
}

@article{lombardi2023statistical,
  title={Statistical modeling of adaptive neural networks explains co-existence of avalanches and oscillations in resting human brain},
  author={Lombardi, Fabrizio and Pepi{\'c}, Selver and Shriki, Oren and Tka{\v{c}}ik, Ga{\v{s}}per and De Martino, Daniele},
  journal={Nature Computational Science},
  volume={3},
  number={3},
  pages={254--263},
  year={2023},
  publisher={Nature Publishing Group US New York}
}

@book{Murray2002,
  author    = {Murray, James D.},
  title     = {Mathematical Biology {I}: An Introduction},
  edition   = {3rd},
  publisher = {Springer},
  year      = {2002},
  doi       = {10.1007/b98868}
}

@article{leone2002ferromagnetic,
  title={Ferromagnetic ordering in graphs with arbitrary degree distribution},
  author={Leone, M and V{\'a}zquez, A and Vespignani, A and Zecchina, Riccardo},
  journal={The European Physical Journal B-Condensed Matter and Complex Systems},
  volume={28},
  number={2},
  pages={191--197},
  year={2002},
  publisher={Springer}
}

@Article{fl2012,
  author = 	 {F. Lombardi and H. J. Herrmann and C. Perrone-Capano and D. Plenz and L. de Arcangelis},
  title = 	 {Balance between excitation and inhibition controls the temporal 
organization of neuronal avalanches},
  journal = 	 {Phys. Rev. Lett},
  year = 	 {2012},
  volume = 	 {108},
  number = 	 {},
  pages = 	 {228703},
}

@Article{lda:06,
  author = 	 {de Arcangelis, L. and Perrone-Capano, C. and Herrmann, H.J.},
  title = 	 {Self-Organized Criticality model for brain plasticity},
  journal = 	 {Phys. Rev. Lett},
  year = 	 {2006},
  volume = 	 {96},
  number = 	 {},
  pages = 	 {028107},
}

@Article{fl2017chaos,
  author = 	 {Fabrizio Lombardi and Hans J. Herrmann  and Lucilla de Arcangelis},
  title = 	 {Balance between excitation and inhibition determines 1/f power spectrum in neuronal networks},
  journal = 	 {Chaos},
  year = 	 {2017},
  volume = 	 {27},
  number = 	 {4},
  pages = 	 {047402},
}

@Article{borgers2005ei,
  author = 	 {Christoph B{\"o}rgers and Nancy J. Kopell },
  title = 	 {Effects of noisy drive on rhythms in networks of excitatory and inhibitory neurons},
  journal = 	 {Neural Computation},
  year = 	 {2005},
  OPTnote = {},
  OPTkey = 	 {},
  volume = 	 {17},
  number = {3},
  pages = 	 {557 - 608},
  OPTmonth = 	 {},
  OPTnote = 	 {},
  OPTannote = 	 {}
}

@Article{wilsoncowan1972,
  author = 	 {Hugh R. Wilson and Jack D. Cowan},
  title = 	 {Excitatory and inhibitory interactions in localized populations of model neurons},
  journal = 	 {Biophysical Journal},
  year = 	 {1972},
  volume = 	 {12},
  number = 	 {},
  pages = 	 {1-24},
}

@article{Buendia_2022_sync,
    author = {Buendía, Victor and Villegas, Pablo and Burioni, Raffaella and Muñoz, Miguel A.},
    title = {The broad edge of synchronization: Griffiths effects and collective phenomena in brain networks},
    journal = {Philosophical Transactions of the Royal Society A: Mathematical, Physical and Engineering Sciences},
    volume = {380},
    number = {2227},
    pages = {20200424},
    year = {2022},
    month = {05},
    issn = {1364-503X},
    doi = {10.1098/rsta.2020.0424},
    url = {https://doi.org/10.1098/rsta.2020.0424},
    eprint = {https://royalsocietypublishing.org/rsta/article-pdf/doi/10.1098/rsta.2020.0424/1325362/rsta.2020.0424.pdf},
}

\end{document}